\documentclass[runningheads]{article}
\usepackage{graphicx}
\usepackage{amsmath,amssymb} % define this before the line numbering.

\usepackage{abstract}
\usepackage{color}
\usepackage[width=122mm,left=12mm,paperwidth=146mm,height=193mm,top=12mm,paperheight=217mm]{geometry}
\begin{document}

\title{Comparison of projection domain, image domain, and comprehensive deep learning for sparse-view X-ray CT image reconstruction} % Replace with your title

\author{Kaichao Liang, Hongkai Yang, Yuxiang Xing\\
Department of Engineering Physics, Tsinghua University}
\maketitle

\begin{abstract}
X-ray Computed Tomography (CT) imaging has been widely used in clinical diagnosis, non-destructive examination, and public safety inspection. Sparse-view (sparse view) CT has great potential in radiation dose reduction and scan acceleration. However, sparse view CT data is insufficient and traditional reconstruction results in severe streaking artifacts. In this work, based on deep learning, we compared image reconstruction performance for  sparse view CT reconstruction with projection domain network, image domain network, and comprehensive network combining projection and image domains. Our study is executed with numerical simulated projection of CT images from real scans. Results demonstrated deep learning networks can effectively reconstruct rich high frequency structural information without streaking artefact commonly seen in sparse view CT. A comprehensive network combining deep learning in both projection domain and image domain can get best results. 
\end{abstract}

\section{Introduction}
X-ray Computed Tomography (CT) is widely used in clinic diagnosis, public safety inspection, non-destructive testing. With complete data acquired, cross-sectional images of scanned objects can be reconstructed from CT data by analytical methods such as filtered back-projection (FBP) \cite{b1} for 2D and FDK \cite{b2} for 3D. In these days, lowering radiation dose \cite{b3,b4} and speedup CT scans \cite{b5} are major concerns in this field. Reducing the number of views in a scan, i.e. sparse view (sparse view) CT, is one of the hot topic currently. With sparse views, acquired data are dramatically insufficient so that analytical reconstruction algorithms result in severe artifacts. In order to achieve good reconstruction, optimization-based iterative methods which can incorporate models of imaging physics, priori information and additional constraints \cite{b6,b7,b8,b9,b10} have gained a lot of attention. For example, studies have shown that good-quality CT image can be reconstructed via constraints of TV minimization \cite{b11,b12,b13}, low rank \cite{b14}, and dictionary learning \cite{b15} from sparse view projection. However, optimization-based iterative reconstruction is computational expensive.

In these years, deep learning has gained a lot of success in many areas such as computer vision, natural language processing, and etc. Specifically, convolutional neural network (CNN) is widely used in image/video processing areas. In the CT field, researchers have done studies using CNN to remove image artifacts for sparse view CT and limited-angle CT \cite{b16,b17,b18} \cite{b20}. Zhu et al proposed an AUTOMAP \cite{b21} which complete domain transform manifold learning for MRI imaging. Considering the imaging modality of CT scan, we examine the capability of deep learning network in catching the features in projection domain and image domain.

\section{Methods and Materials}

\subsection{Physics Model of X-ray CT Imaging}
In X-ray CT, photons emitted from an X-ray source are attenuated by the object to be imaged along its ray path according to Beer's law. Normally, measurement process is modeled as:

\begin{align}
\label{Eq:monoApprox}
  p(l) & = -\ln\frac{\int_{E}\emph{S}(E)e^{-\int_{E}\mu(E,\textbf{r})d\textbf{r}}dE + \mbox{noise}}{\int_{E}S(E)dE} \cong \int_{l}\widetilde{\mu}(\textbf{r})d\textbf{r}+ \mbox{noise}
\end{align}
with $S(E)$ the energy spectrum of the system with detector response accounted, $\mu(E,\mathbf{r})$ the linear attenuation map of a scanned object at location $\mathbf{r}$, and $l$ the ray path from the source to a detector bin. Here, the $\widetilde{\mu}$ denoting the effective attenuation coefficient under monochromatic asumption.

For a practical scan, we denote projection data as a vector $\mathbf{p} \in \mathbb{R}^{M \times 1} $ with $M=VD$ in case of $D$ detector bins and $V$ projection views. Hence, in discrete form, Eq.~(\ref{Eq:monoApprox}) can be written as:
\begin{align}
\label{Eq:discretePrjModel}
  \textbf{p} &= \textbf{H}\boldsymbol{\mu}+\textbf{n}
\end{align}
where $\boldsymbol{\mu} \in \mathbb{R}^{N \times 1}$ is a vector denoting a $\sqrt{N} \times \sqrt{N}$ image of effective attenuation map which is the discretized image $\widetilde{\mu}(\mathbf{r})$ within the field of view. The $\mathbf{H} \in \mathbb{R}^{M \times N}$ is the system matrix with its elements $H_{ij}$ descripting the contribution from pixel $j$ to $i^{\rm{th}}$ ray path, and $\mathbf{n}$ is a zero-mean noise vector. One can model more physics in $\mathbf{H}$ though we only consider the basic line integration model of X-ray CT imaging here. Reconstructing $\boldsymbol{\mu}$ is to solve Eq.~(\ref{Eq:discretePrjModel}) in general.

In sparse view case, projection views is reduced by a significant amount to be $V^{\rm{sp}} \ll V$ with the superscript $^{\rm{sp}}$ short for sparse. Hence, only $\mathbf{p}^{\rm{sp}} \in \mathbb{R}^{DV^{\rm{sp}} \times 1}$ is available and the problem in Eq.~(\ref{Eq:discretePrjModel}) is now severely ill-posed. We denote the sparsity factor to be $B = \frac{V}{V^{\rm{sp}}}\gg 1$. Optimization under the Bayesian framework is commonly used to solve such a problem. However, it is time consuming. To facilitate the sparse view CT reconstruction and get a reconstruction of reasonable image quality, deep learning can be useful.

\subsection{Architectures of the Networks}
\begin{figure}[tbp]
   \centering

   \includegraphics[width=0.96\linewidth]{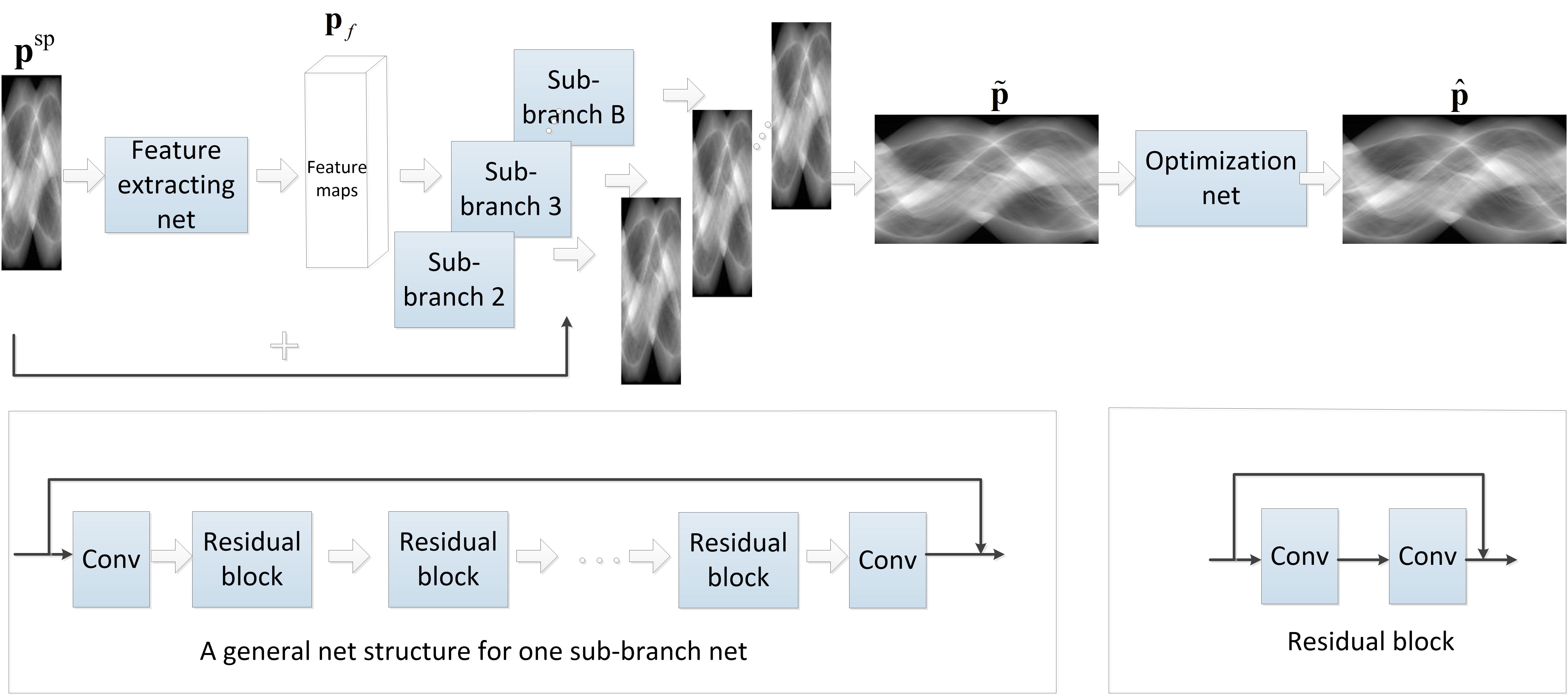}
\caption{The architecture of projection estimation network. The block A is one sub-branch of the projection estimation network which serves to estimate one subset of projection. The block B is the basic constituent module which is composed of two convolution layers plus batch norm and the ReLu activation. It is with a residual connection same as the ResNet\cite{b22} }
\label{fig:prjEstNet}
\end{figure}

\begin{figure}[tbp]
   \centering

   \includegraphics[width=0.8\linewidth]{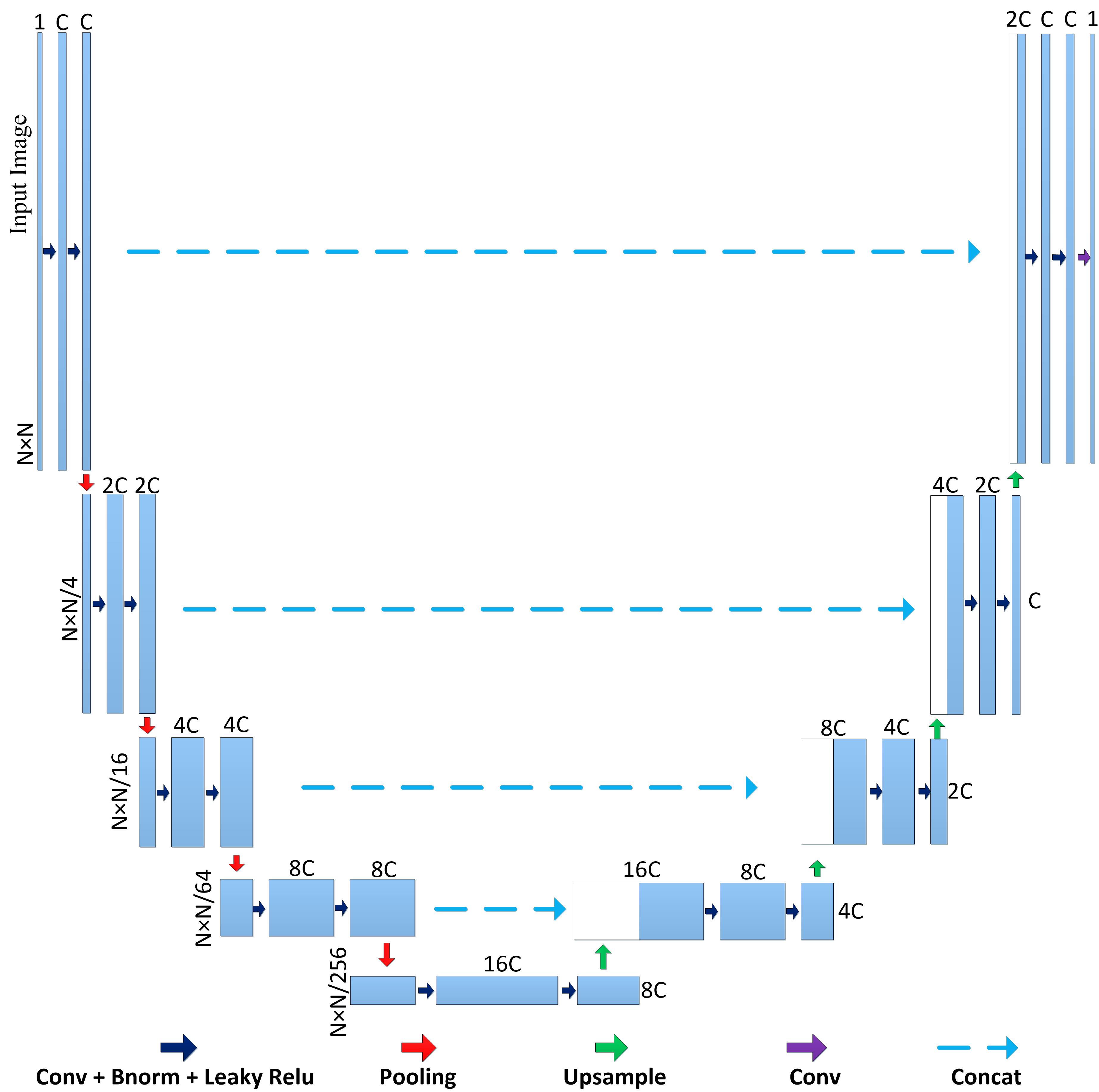}
\caption{The architecture of U-net in image domain for sparse view CT reconstruction. }
\label{fig:Unet}
\end{figure}

\begin{figure}[tbp]
   \centering

   \includegraphics[width=0.96\linewidth]{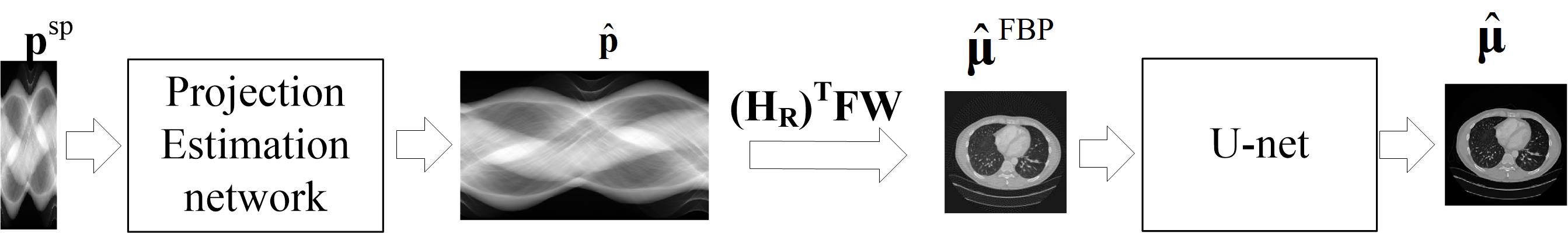}
\caption{The architecture of comprehensive network combining projection estimation and U-net in image domain for sparse view CT reconstruction. }
\label{fig:overallNet}
\end{figure}

Here we compare three types of deep learning network architectures: 1) Estimate missing projections by deep learning and reconstruction by FBP afterwards. 2) Reconstruction with FBP using sparse-view data and using deep learning network to reduce artefacts; 3) Network estimating missing projections + FBP + image domain network. The architectures of these networks are shown in Fig.~\ref{fig:prjEstNet}, ~\ref{fig:Unet}, and \ref{fig:overallNet} respectively.

The projection estimation network is to estimate missing projection data from sparse view projections $\mathbf{p}^{\rm{sp}}$ to gain complete data \cite{b23}.
As Fig. ~\ref{fig:prjEstNet} shows, the projection estimation network is composed of three functional blocks.
The first block of net is feature extracting net. It takes $\mathbf{p}^{\rm{sp}}$   as input and extracts input projection map's features with convolutional layers. Defining the function of feature extracting block as $\phi_{\rm{f}}$, the extracted feature maps of three dimension can be denoted as: $\mathbf{p}_{\rm{f}} = \phi_{\rm{f}}(\mathbf{p}^{\rm{sp}})$.
The second block is consist of projection estimation sub-branches. According to the CT scanning geometry, missing views are grouped into subsets according to their neighborhood relationship,e.g., B subsets in case of sparsity factor being B. Considering different projection subsets have different angular distance from the acquired projection $\mathbf{p}^{\rm{sp}}$, $B-1$ sub-branch nets are designed to estimate the  $B-1$ subsets of projections separately \footnote{The $\mathbf{p}^{\rm{sp}}$ are one of the subsets which is known}. These nets take feature maps as inputs and finally generate  $\Delta\mathbf{p}_1,  \Delta\mathbf{p}_2, ..., \Delta \mathbf{p}_B$ to estimate $\widetilde{\mathbf{p}}_1, \widetilde{\mathbf{p}}_2, \dots, \widetilde{\mathbf{p}}_B$. Defining the estimation operator as $\varphi_b$  , then  $\widetilde{\mathbf{p}}_b =  \Delta \mathbf{p}_b+\mathbf{p}^{\rm{sp}}=\varphi_b(\mathbf{p}_{\rm{f}})+\mathbf{p}^{\rm{sp}}$.
At the end of prj-estimating sub-branches, the estimated projections together with input sparse-view projection are concatenated in the order of view angles to form full-view projections  $\widetilde{\mathbf{p}} = \rm{concat}(\widetilde{\mathbf{p}}_1,\widetilde{\mathbf{p}}_2, \dots, \widetilde{\mathbf{p}}_B)$.
A third block is added to unify the projections coming from different sub-branches so that the full-view projection are further tuned to be consistent as a complete dataset. The final output of full-view projection estimated is denoted as $\widehat{\mathbf{p}}$ . The main structure of this network is consist of residual blocks \cite{b22}. All convolutional layers are followed with leaky-relu activation and batch-norm layers \cite{b24}, except for the output layers of branches and last output layer. The output of this network is reconstructed by FBP to get reconstruction images.

Applying U-net in image domain for sparse-view CT is very straightfoward. The U-net structure used in this study is as shown in Fig. ~\ref{fig:Unet}. This network takes FBP reconstruction of sparse-view data as input.

For comprehensive network, as shown in Fig. ~\ref{fig:overallNet}, a sinogram-to-image domain transform is configured as a layer in the network. With this layer built in, one is able to trace the error propagation in a comprehensive way. This layer mainly complete three computations: weighting, filtration and weighted back-projection. This can be realized by matrix-vector multiplication:
\begin{align}
\label{Eq:FBPrecon}
  \widehat{\boldsymbol{\mu}}^{\rm{FBP}} &= \mathbf{H}_{\rm{R}}^{\rm{T}} \mathbf{FW}\widehat{\mathbf{p}}
\end{align}
Here, $\mathbf{W}$ is a diagonal matrix, the matrix $\mathbf{F}$ completes a ramp filtration in detector axis for all views, and $\mathbf{H}^{\rm{T}}_{\rm{R}}$ the back-projection operator with the superscript T denoting matrix transform. Please notice that $\mathbf{H}_{\rm{R}}$ differs from the  $\mathbf{H}$ in Eq. (3) because additional weighting would be incorporated in  $\mathbf{H}^{\rm{T}}_{\rm{R}}$  in case of fan-beam and cone-beam CT scan according to analytical reconstruction methods. These three matrices are predetermined by the CT scanning geometry and can be pre-calculated. Obviously, it is easy to accommodate more physical factors in these matrices.
This comprehensive network outputs the final reconstruction.

\subsection{Network Training}

Projestion estimation network is trained using an $l_2$-norm loss function:

\begin{align}
  \rm{argmin}\sum_{k=1}^{K}||\widehat{\mathbf{p}}-\mathbf{p}||_{2}^{2}
\end{align}
The true complete dataset $\mathbf{p}$  are used as the labels.
The image domain U-net and comprehensive network are trained by ultimate loss of reconstruction error of $l_2$ -norm:
\begin{align}
\varepsilon = \frac{1}{K} \sum_{k=1}^{K} \| \widehat{\boldsymbol{\mu}}_k - \boldsymbol{\mu}_k \| _{2}^{2}
\end{align}
with $\boldsymbol{\mu}_{k}$ being the known label, and $K$ being the number of images in a training set.

To examine the performance of these networks, we conduct our study on reconstructing a sparse view 2D fan-beam CT. Projections of evenly distributed views over $2\pi$ are simulated to acquire sinogram data. Projections at 360 views are considered complete dataset aimed in the projection estimation network. Sparse view data are draw from the complete data according to a sparse factor.

In total, 17720 thoracic CT images from real scans of 52 patients \cite{b25} were used as phantoms in our study. Among them, 15720 phantom images from 48 patients were used as training set, while 2000 phantom images from non-overlapping 4 independent patients were used as test set (ground truth only used for performance evaluation). Fan-beam CT scan was simulated by homemade simulation toolkit to gain projection data. For the scanning geometry, both source-to-origin and detector-to-origin distances were 1000 mm. A linear detector of 216 bins with bin size 5 mm was applied. Considering the memory limitation of our GPU server, the reconstruction area was on the 200 grid with pixel size 2.56 $\mbox{mm}^{2}$. 360-view simulated projections were used as labels in the training of the projection estimation network. The truth of phantom images were used as labels in the loss of image domain U-net and comprehensive network. We trained the networks for the cases of 72-views, $B=5$ and 45-views, $B = 8$. In training, the network is able to reach reasonable performance before 10 epochs.

\section{Results}
\begin{figure}[tbp]

   \centering
   \includegraphics[width=0.95\linewidth]{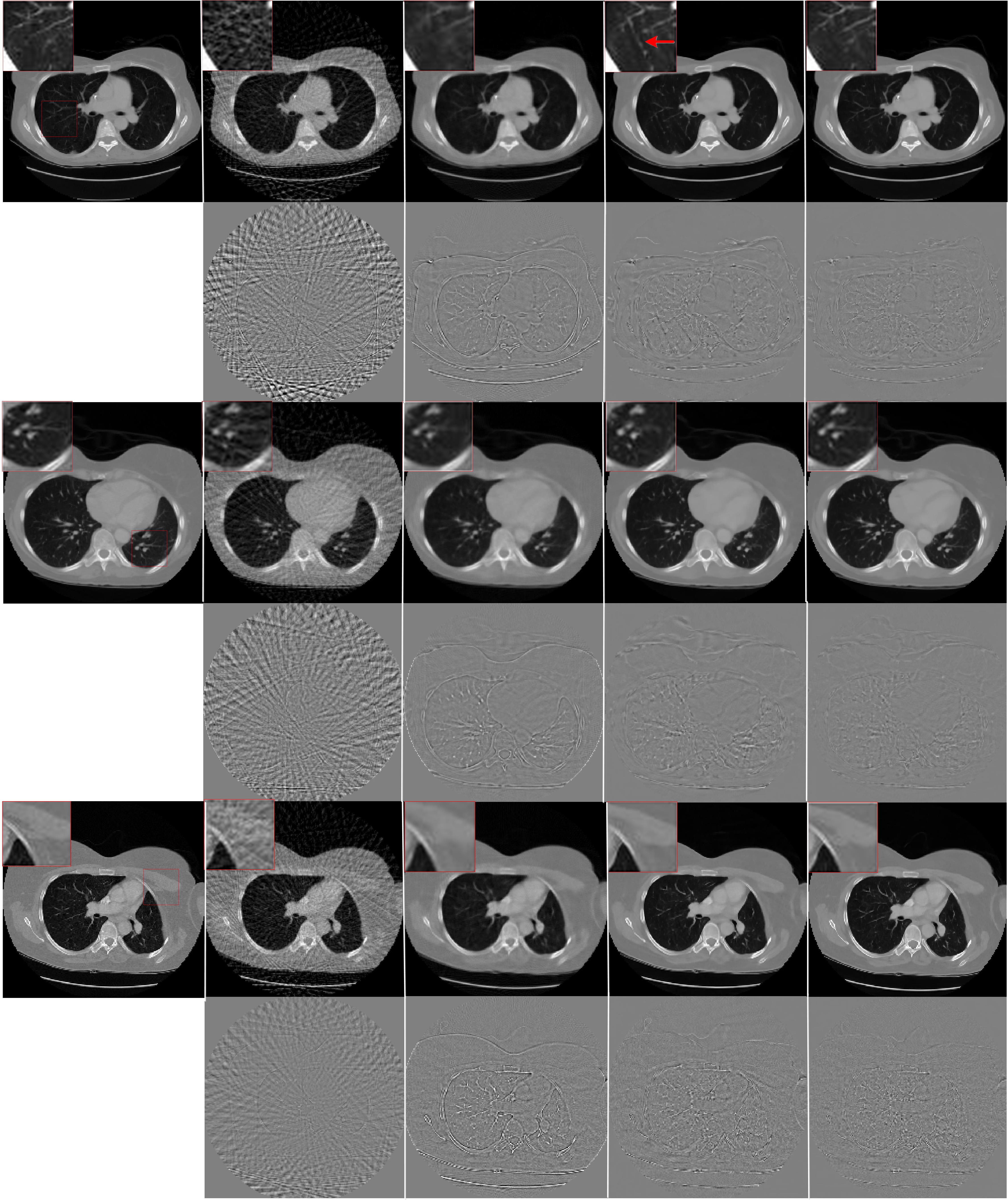}
\caption{Three cases of reconstructions. From left to right: phantoms, 72-veiw FBP, 72-view projection estimation reconstruction, 72-view U-Net, and the proposed network reconstructions. In each case, details in the red boxes are displayed on upper left corner. The differences of the four reconstructions of 72-view data from the phantom are displayed below the corresponding reconstructions. Images in a same category are in same gray scale}
\label{Fig:72ViewRecon}
\end{figure}

\begin{figure}[tbp]

   \centering

   \includegraphics[width=0.95\linewidth]{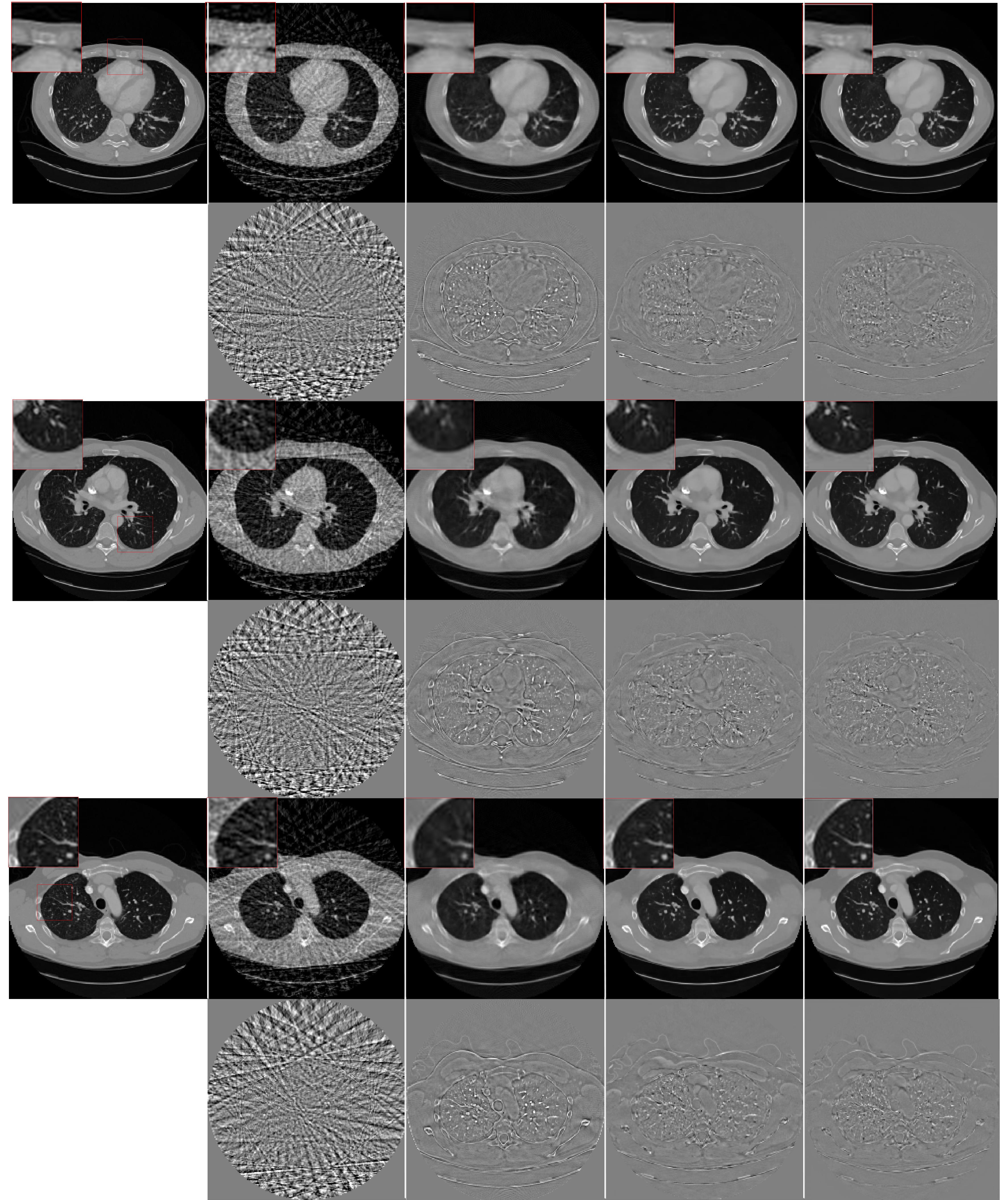}
\caption{Three cases of reconstructions. From left to right: phantoms, 45-veiw FBP, 45-view projection estimation reconstruction, 45-view U-Net, and the proposed network reconstructions. In each case, details in the red boxes are displayed on upper left corner. The differences of the four reconstructions of 45-view data from the phantom are displayed below the corresponding reconstructions. Images in a same category are in same gray scale}
\label{Fig:45ViewRecon}
\end{figure}

Using the trained network, simulated sparse-view CT projections of test phantoms are reconstructed by feeding sparse-view projections to the input for the projection estimation network and comprehensive network. FBP reconstructions are fed into U-net.  Fig.~\ref{Fig:72ViewRecon} is three cases of 72-view results from test set and Fig.~\ref{Fig:45ViewRecon} is three cases of 45-view results from test set. Sparse-view FBP results and ground-truth phantoms are also displayed for an overall comparison. In the upper left of reconstruction images, we also display zoom-in of a regions of interest (ROI) (indicated by the red squre boxes for details. Difference images from the truth are also displayed below the reconstructed images. We can see that FBP reconstructions is severely contaminated by streaking artefacts. Though projection estimation network removed streaking artefacts by estimating the missing data, the images were still blurry with some details lost. Image-domain U-Net produced visually pleasant images compared with projection estimation network as the loss-function is directly formed in image-domain. However, fake structures might show up (as indicated by the arrow in Fig. \ref{Fig:72ViewRecon} in the lung). We think this is because some  streaking artefacts are similar to structures. The comprehensive network achieved more precise reconstruction than both projection-domain network and image-domain network. Streaking artefacts were removed and details were well preserved.

Moreover, we quantitatively evaluated the image quality of all test reconstructions in terms of relative root mean square error (RRMSE) and the structural similarity (SSIM) index:

\begin{align}
  \rm{RRMSE} &= \frac{||\widehat{\boldsymbol{\mu}}-\boldsymbol{\mu}||_{2}}{||\boldsymbol{\mu}||_{2}}
\end{align}
\begin{align}
  \rm{SSIM} &= \frac{(2\bar{\widehat{\boldsymbol{\mu}}}\bar{\boldsymbol{\mu}}+C_{1})(2\sigma_{\widehat{\boldsymbol{\mu}}\boldsymbol{\mu}}+C_{2}))}{{(\bar{\widehat{\boldsymbol{\mu}}}^{2}+\bar{\boldsymbol{\mu}}^{2}+C_{1})(\sigma_{\widehat{\boldsymbol{\mu}}}^{2}+\sigma_{\boldsymbol{\mu}}^2+C_{2})}}
\end{align}
with  $\widehat{\boldsymbol{\mu}}$ being a reconstruction and $\boldsymbol{\mu}$  the corresponding label image, $\bar{\widehat{\boldsymbol{\mu}}}$  and $\bar{\boldsymbol{\mu}}$  the means of  $\widehat{\boldsymbol{\mu}}$ and $\boldsymbol{\mu}$  respectively,  $\sigma_{\widehat{\boldsymbol{\mu}}}$ and $\sigma_{\boldsymbol{\mu}}$  the  standard deviations, and $\sigma_{\widehat{\boldsymbol{\mu}}\boldsymbol{\mu}}$  the cross-correlation. The constants $C_{1}$ and $C_{2}$ are stabilizers.

We calculated the ensemble means and standard deviations of image-by-image RRMSE and SSIM for performance evaluation over the whole testing dataset. Results are in Table \ref{Table:meanStd}. The comprehensive network achieved an overall RRMSE (include pixels in all images) 0.0474 for the 72-view test case and 0.0677 for the 45-view test case. These figures were significantly improved from FBP results indicating the effectiveness of deep learning networks. The comprehensive structure achieved better RRMSE and SSIM results than both projection-domain and image-domain network. This is due to the complementarity of informations in image-domain and projection-domain. The comprehensive network also had an advantage on stability as the stds of RRMSE and SSIM  over the whole test set were the smallest among the listed methods.

\begin{table}[htp]
\centering
\caption{Means and Stds of RRMSE and SSIM over the whole test dataset.}
\label{Table:meanStd}
\begin{tabular}{ccc}
\hline
 Methods & RRMSE & SSIM \\
\hline
  72-view FBP & $0.1538 \pm 0.031$ & $0.9786 \pm 0.0076$\\
  72-view Prj Estimation network & $0.0762 \pm 0.021$ & $0.9944 \pm 0.0031$\\
  72-view U-Net & $0.0575 \pm 0.017$ & $0.9968 \pm 0.0018$\\
  72-view comprehensive network & $\mathbf{0.0474 \pm 0.015}$ & $\mathbf{0.9978\pm 0.0014}$\\
  45-view FBP & $0.2667 \pm 0.033$ & $0.9457 \pm 0.0107$\\
  45-view Prj Estimation & $0.1062 \pm 0.023$ & $0.9906 \pm 0.0037$\\
  45-view U-Net & $0.0790 \pm 0.018$ & $0.9948 \pm 0.0022$\\
  45-view comprehensive network & $\mathbf{0.0677 \pm 0.016}$ & $\mathbf{0.9962\pm 0.0019}$\\

\hline
\end{tabular}
\end{table}

\section{Discussion and Conclusions}

We compared three strategies of reconstruction based on deep learning for sparse view CT in this work. The projection estimation network tested is of a Res-CNN structure. The image domain network takes the advantage of the good performance of U-net. The comprehensive network combined projection estimation deep learning, analytical inversion transform, and image domain leaning respectively. Our experimental results with realistic thoracic phantoms show that all of the networks performed quite nicely for sparse view CT reconstruction. All reconstructions are free of streaking artefacts which is often a big problem in a sparse view CT. The comprehensive network wins by integrating the power of deep learning in both projection and image domains. The end-to-end training is made possible by incorporating the analytical FBP operator in as a network layer. Its performance is better than recovering information in projection domain or image domain only. The overall $l_2$ loss in images propagate through the projection model of CT imaging to update weights in projection estimation network. Though the comparison in this work is somehow unfair because the comprehensive network is bigger and involves more computation, we still think its meaningful to illustrate the effects of deep learning in two different domains for a complex problem like CT. We would further our efforts to explore more networks architectures and hyper-parameter optimization for CT reconstruction in our future work. 

These deep learning reconstruction networks can be easily extended to 3D imaging, as well other imaging modalities such as MRI, PET and SPECT imaging.

%\clearpage

\bibliographystyle{splncs}

\end{document}